\documentclass[aps,a4paper,12pt]{article}
\usepackage{latexsym}
\usepackage{amsmath,amssymb}
\usepackage{graphics}
\usepackage{epsfig}
\def\be{\begin{equation}}
\def\fin{\end{equation}}
\def\disp{\displaystyle}

\def\scri{\scriptsize}
\def\h{{\cal H}}
\def\ho{Harper-Hofstadter   }

\def\T{{\sf T\kern-.45em T}}
\def\C{\kern.1em{\raise.47ex\hbox{$\scriptscriptstyle |$}}
             \kern-.40em{\sf C}}
\def\Z{{\mathbb{Z}}}

\def\al{\alpha}
\def\slz{PSL(2,{\mathbb{Z}})}
\def\slr{PSL(2,{\mathbb{R}})}
\def\hfl{\disp\mathop{\hbox to 10mm{\rightarrowfill}}}

\def\qed{$\ \Box$}
\begin{document}
\title{Random Walks on the Braid Group $B_3$ and Magnetic Translations in Hyperbolic Geometry}
\author{Rapha\"el Voituriez\\
{\it Laboratoire de Physique Th\'eorique et Mod\`eles
Statistiques},\\ {\it Universit\'e Paris Sud, 91405 Orsay Cedex, France.}\\ email: {\it voiturie@ipno.in2p3.fr}}

\maketitle
\begin{abstract}
We study random walks on the three-strand braid group $B_3$, and in particular compute the drift, or average topological complexity of a random braid, as well as the probability of trivial entanglement. These results involve the study of magnetic random walks on hyperbolic graphs (hyperbolic \ho  problem), what enables to build a faithful representation of $B_3$ as generalized magnetic translation operators for the problem of a quantum particle on the hyperbolic plane.\end{abstract}

{\bf PACS:} 02.50.Ga, 03.65.Fd, 02.20.Bb, 02.20.Rt

{\bf Key words:} braid groups, discrete magnetic Shr\"odinger operators, representation theory, hyperbolic geometry.
\section{Introduction and basic definitions}
\label{sect1}
This paper presents different results concerning random walks on the three--strand braid group  $B_3$. Besides the relevance of this work at a purely algebraic level (see \cite{birman} for a review of knot theory), its usefulness in the understanding of topological problems in statistical physics is also undeniable. We will keep in mind the picture of entanglements of chain--like objects, like polymers \cite{nech_gros} or vortex lines in super conductors \cite{nelson}. The tools involved in this work unexpectedly lead to the study of the quantum mechanics of a charged particle moving on the hyperbolic plane in a constant magnetic field,  and allow a discrete interpretation of an exact result  \cite{matsumoto} giving the probability distribution of the area enclosed by a hyperbolic Brownian motion. This connection between  $B_3$ and quantum mechanics in hyperbolic geometry is the main new result of this paper; it is established by proposition 2 and theorem 1.  The mathematical ingredients of this theorem, such as group cohomology and central extensions of groups, as well as the concept of generalized magnetic translation operators of definition 3,  are introduced in a general way, in order to show that this interpretation of a central extension as a magnetic field is a general result that can be adapted to other extensions of lattice groups. This paper focuses on the case of $B_3$, because of its physical importance; in particular, using the relation between $B_3$ and $\slz$ and between the corresponding $C^{\star}$--algebras enlightened by proposition 3, we obtain  theorem 2 and theorem 3,  which provide  rigorous derivations of the return probability and the drift for $B_3$, results which were more intuitively given in \cite{voitnech3}. Here we use the more systematic method of studying the transition operator and the algebra it belongs to, which allows a possible  generalization of these results to other Abelian extensions of hyperbolic groups.

\begin{figure}[ht]
\begin{center}
\epsfig{file=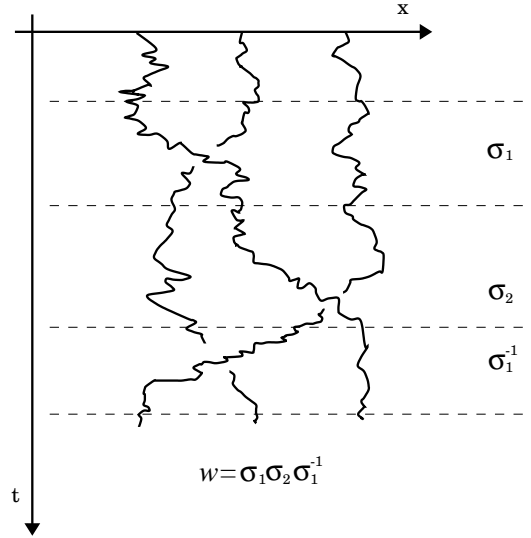,width=7cm}
\end{center}
\caption{Braiding of directed polymers. Elementary moves $\sigma_i$ generate $B_3$.}
\label{fig:1}
\end{figure}

Consider the 3D model of $N$ directed polymer chains, labelled by $\vec{r}_{i}(t)=(x_i(t),y_i(t))$,  attached at one end (on the plane $t=0$), growing along the direction $t>0$ and fluctuating in the plane  $(x,y)$. We assume periodic boundary condition, that is $\{\vec{r}_{i}(T)\}=\{\vec{r}_{p(i)}(0)\})$, $p$ being a permutation over $N$ elements. We are interested in the complexity of this randomly generated entanglement, which can play a crucial role in macroscopic physical properties, such as polymer elasticity.  Each topological configuration can be exactly encoded by an element of $B_N$, that is a word whose letters are the generators of the group, associated to elementary moves as shown in figure \ref{fig:1}. Looking at the projection onto the plane $(t,x)$, elementary moves correspond to positive/negative crossings.  We restrict our study to the simplest non commutative case $N=3$, whose set of generators is  $S=\{\sigma_1,\sigma_2, \sigma_1^{-1},\sigma_2^{-1}\}$. A randomly generated configuration corresponds to an element $w_n$,  random product of $n$ generators; in other terms we focus on the following Markov process
\be
\begin{array}{ll}
w_{0}={\rm Id}\\
w_{n}=w_{n-1}\zeta_n\ {\rm for}\ n\ge1
\end{array}
\end{equation}
where $\zeta_n$ are i.i.d. random variables in $S$. Average over this random  variable will be denoted by $\left<.\right>$. A more thorough study would require to establish a law $n(T)$, giving the time scaling of the number of elementary moves,  which is not discussed here. We study the probability distribution of $L(w_n)$, the irreducible length in the metric of words, or in other terms the minimal number of generators necessary to build $w_n$. $L(w_n)$ is exactly the minimal number of crossings necessary to represent the entanglement, and can therefore be seen as an indicator of the braid complexity. In particular, if $L(w_n)=0$ the braid is trivial (or unlinked). 

We now introduce $B_3$ in a more algebraic way, useful to study random processes on this group. 

\vspace{0.8cm}
{\bf Definition 1.} {\it $B_3$ is defined by the set of generators $S=\{\sigma_1,\sigma_2, \sigma_1^{-1},\sigma_2^{-1}\}$ together with the commutation relation
\be\label{artin}
\sigma_1\sigma_2\sigma_1=\sigma_2\sigma_1\sigma_2
\end{equation}
or equivalently by the generators $S'=\{a,b,a^{-1},b^{-1}\}$ together with the relation}
\be
a^2=b^3
\end{equation}
The equivalence of these (so-called Artin) presentations is explicit if one sets:
\be
\begin{array}{l}\label{eq2}
a=\sigma_1\sigma_2\sigma_1\\
b=\sigma_1\sigma_2
\end{array}
\end{equation}
It is known that $B_3$ is a central extension of $PSL(2,\Z)$\footnote{the group of $2\times2$ unimodular matrices of integer entries, factored by $\pm{\rm Id}$}, i--e that $B_3$ factored by its center $Z$ is isomorphic to $\slz$. Let us recall that the center $Z$ of $B_3$, which is the subgroup of elements commuting with any element of $B_3$, is  generated by $a^2=b^3=\Delta^{2}$. $\Delta^2$ is the usual notation for the generator of the center in Artin groups theory. For $B_3$,  one has simply $\Delta=a$. It is useful to introduce 
\be
i:\left\{\begin{array}{ccc}
\Z&\longrightarrow& Z\\
n& \longrightarrow& i(n)=\Delta^{2n}
\end{array}\right.
\end{equation}
which shows that  $Z$ is isomorphic to $\Z$. We then denote by $\pi$ the
canonical projection onto $\slz$  and by $s$ a cross--section of $\pi$, that is an application such that $\pi\circ s={\rm id}$. It is convenient to  summarize these elements by writing the following exact sequence:
\be\label{seq}
0 \hfl^{}_{} \Z  \hfl^{i}_{} B_3  \hfl^{\pi}_{} \slz \hfl^{}_{} 1 
\end{equation} 
Note that $\slz$ is a free product $\Z_2\star\Z_3$, $\pi(a)$ and $\pi(b)$ being the corresponding generators of order 2 and 3 respectively.
The generators of the group $B_3$ can be represented by 
$GL(2,\Z[u^{\pm1}])$--matrices. To be more precise:

\vspace{0.8cm}{\bf Definition 2.} (see \cite{birman})  {\it The Burau representation of $B_3$ is given by} 
\be\label{eq:4}\sigma_1=\left(\begin{array}{cc} u^2 & 1 \\ 0 & 1
\end{array}\right); \qquad
\sigma_2=\left(\begin{array}{cc} 1 & 0 \\ -u^2 & u^2
\end{array}\right)
\end{equation}
{\it where $u$ is a free parameter. This representation is faithful.}

We conveniently introduce an auxiliary group,  $\slz_{u}$, built on generators (\ref{eq2}) of this representation of $B_3$ normalized by the determinant:

\be \label{eq:4bis}
a_u=\left(\begin{array}{cc} 0 & 1/u \\ -u & 0
\end{array}\right); \qquad
b_{u}=\left(\begin{array}{cc} 0 & 1 \\ -1 & 1
\end{array}\right)
\end{equation}
This group is a ``deformation'' of $\slz$, which preserves all commutation relations. For $u=1$, one has $\slz_{u}=\slz$.  
 For a more detailed discussion of braid
groups the reader is referred to \cite{dehornoy,birman}.

\section{Random processes on $B_3$: magnetic random walk formalism}

\begin{figure}[ht]
\begin{center}
\epsfig{file=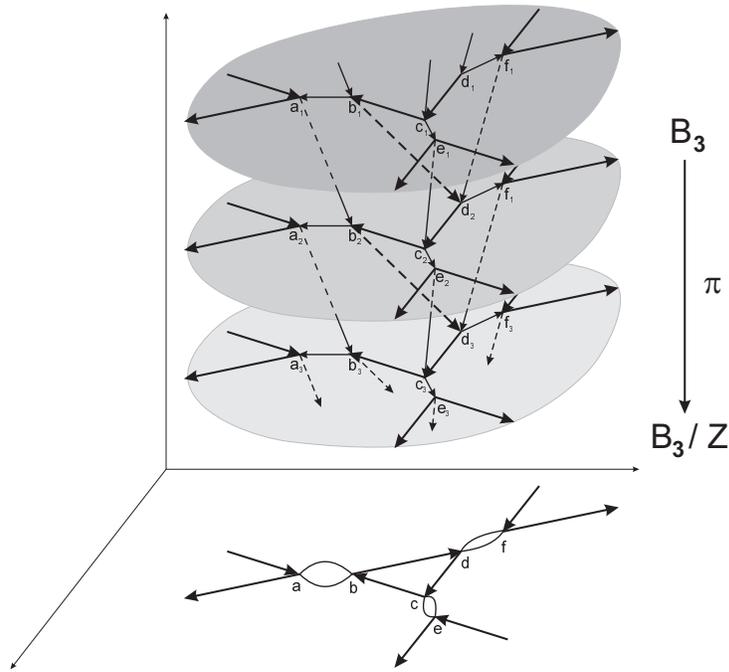,width=7cm}
\end{center}
\bigskip

\caption{$B_3$ Cayley graph and its projection ($\slz$). Thin arrows correspond
to 
$a$, thick ones to $b$. Note that $\pi(\alpha_i)=\alpha$,  
$\pi(\beta_i)=\beta$ and so on. Recall that $\pi(a)$ has to be identified with 
$\pi(a)^{-1}$}
\label{fig:4}
\end{figure}

We now show that a free random walk  on $B_3$ can be viewed as a random walk on $PSL(2,\Z)$ with magnetic field (we will say a magnetic random walk). The key point  is that $B_3$  is a central extension of 
$PSL(2,\Z)$.  We will use systematically the set of generators $S'=\{a,a^{-1},b,b^{-1}\}$, to avoid technical complexity, keeping in mind that the formalism can be adapted to generators $S=\{\sigma_1,\sigma_2, \sigma_1^{-1},\sigma_2^{-1}\}$ introduced previously. Results of physical interest will be given in terms of these generators $S$. First note that the Cayley graph\footnote{The graph with set $B_3$ of vertices, and set $\{\{x,xs\}:x\in B_3,s\in S'\}$ of edges} of $B_3$ is three dimensional.  As shown in Fig.\ref{fig:4}, the map $\pi$ can then be viewed as a projection from 3D to 2D; the vertical axis represents the ``center'' coordinate. Intuitively, we will show that this center term counts the magnetic flux through the random path. 


Namely, if one inserts in each elementary cell of the hyperbolic lattice a unit "magnetic flux" (see Fig.\ref{fig:4}) and denotes by 
$f$ the total flux through a closed path on $\slz$, then any word $w^{Z}_n$ corresponding to a walk ending in $Z(B_3)$ (note that  $\pi(w^{Z}_n)$ is a closed path on $\slz$) can be written as
$$
w^{Z}_n={\Delta}^{2f}
$$ 
Hence the function $p_n(f)$ giving the joint probability to obtain a closed $n$--step 
loop on the  graph $\slz$ carrying a flux $f$ is of great interest, especially 
because $p_n(0)$ is the probability to get a {\it trivial braid} (i.e. completely reducible word).

Let us establish more rigorously this relation between $B_3$ and what could be interpreted as a hyperbolic version of the Harper-Hofstadter problem \cite{harper}. To introduce the usual setting of this problem, we define the transition operator, or discrete Laplace operator,
\be
h=\frac{1}{4}(a+a^{-1}+b+b^{-1})
\end{equation}
viewed as an element of $C^{\star}(B_3)$\footnote{That is the algebra generated by the left regular representation of $B_3$ on $l^2(B_3)$}. We use the following decomposition for any $X\in C^{\star}(B_3)$
\be
X=\sum_{w\in B_3}c_{w}(X)w
\end{equation}
with $c_w\in {\mathbb C}$ and the elements $w$ of $B_3$ are identified with the operators  acting  on $l^2(B_3)$. Different quantities characterizing random walks on $B_3$ are obtained from $h$. One is the return probability
\be
p_n(0)=c_{{\rm Id}}(h^n)\equiv\tau(h^n)
\end{equation}
where Id denotes the identity operator. This scalar component $\tau$  of an operator is called a tracial state on $C^{\star}(B_3)$. Another quantity of interest is the drift
\be\label{drift}
l_{B_3}=\lim_{n\to\infty}\frac{\left<L^{B_3}(w_n)\right>}{n}=\lim_{n\to\infty}\frac{1}{n}\sum_{w\in B_3}c_{w}(h^n)L^{B_3}(w)
\end{equation}
Note that these quantities depend only on the algebraic structure of $C^{\star}(B_3)$; they are purely combinatorial objects which in particular do not depend on the representation of the algebra. We are therefore allowed to use a  more convenient isomorphic algebra, what is the aim of the following construction. Following the definition of rotation algebras in \cite{beguin,rieffel}, we build a   $C^\star$--algebra $R_\theta$, generated by $A_\theta, B_\theta$ defined formally by the following $\star$--homomorphism
\be
\Psi_{\theta}:\left\{
\begin{array}{llll}
a&\longrightarrow&A_\theta\\
b&\longrightarrow&B_\theta\\
\Delta^2&\longrightarrow&e^{i\theta}
\end{array}\right.
\end{equation}
The aim of the following is to construct explicitly a representation of  $R_\theta$ on $l^2(\slz)$, and to show that   $\Psi_{\theta}(h)$ is  exactly the  Hamiltonian of a particle moving on the graph $\slz$ in a magnetic field of elementary flux $\theta$. This point accounts more rigorously for the geometrical picture given above. In a more mathematical language, we use general results of group cohomology (see \cite{brown} for review). Namely, we use the following proposition:

\vspace{0.8cm}{\bf Proposition 1.}\label{iso1}  {\it There exists an integer--valued map $\Theta$  satisfying
\be\label{cocy}
\Theta(x,yz)+\Theta(y,z)=\Theta(xy,z)+\Theta(x,y),
\end{equation}
such that $B_3$ is isomorphic to the Cartesian product $\Z\times\slz$ with the following multiplication
\be\label{mult}
(m,x).(n,y)=(m+n+\Theta(x,y),xy)
\end{equation}}
{\it Proof.} This is a direct application of a general theorem of classification of extensions with Abelian kernel (see \cite{brown}) to the decomposition (\ref{seq}) of $B_3$ .  $\Theta$ is an element of $H^2(\slz,\Z)$,  the set of so--called group 2--cocycles, and is unique up to cohomology equivalence. Associativity of the product (\ref{mult}) is ensured by (\ref{cocy}). \qed

We now trivially map $\Z\times\slz$ into $U(1)\times\slz$ via $(n,w)\rightarrow e^{i n\theta}w$. Now the identification
\be
\begin{array}{lll}
A_\theta & \longleftrightarrow & (0,\pi(a))\\
B_\theta & \longleftrightarrow & (0,\pi(b))
\end{array}
\end{equation}
clarifies our point: $B_3$ (via $R_\theta$) is identified with $U(1)\times\slz$, which is a ``magnetic'' equivalent of $\slz$. Indeed the 2--cocycle $\Theta$ is the discrete analogue of a 2--form, or a magnetic field in physical terms. Cohomology equivalence is then interpreted as  gauge invariance. We believe that this point, put forward in details in \cite{mathai}, is very important and that this magnetic  interpretation of central extensions can be fruitfully generalized to other groups.

\section{Quantum mechanics in hyperbolic geometry}
We now come to the explicit construction of a representation of $R_\theta$ on $l^2(\slz)$, showing in particular that $\Psi_{\theta}(h)$ is  exactly the  Hamiltonian of a particle moving on the graph $\slz$ in a magnetic field of elementary flux $\theta$. Sunada has given in \cite{sunada} a general definition of discrete magnetic Schr\"odinger operators ({\it generalized Harper operators}). We here wish to give a construction compatible with this definition, starting from the continuous problem.      
\subsection{Continuous case}
We consider the quantum mechanics of a charged particle moving on the hyperbolic plane in a constant magnetic field $B$. This problem has been extensively studied in \cite{alain}. Let us recall some results of \cite{alain}. We  denote by $\h$ the upper half-plane of the complex variable $z=x+iy\ (y>0)$ endowed with a  metric $g$ such that
\be
ds^2=\frac{dx^2+dy^2}{y^2}
\end{equation} 
The Lagrangian of the particle, of mass $1/2$ and charge 1, reads
\be
{\cal L}=\frac{{\dot x}^2+{\dot y}^2}{y^2}-B\frac{{\dot x}}{y}
\end{equation}
A thorough study of the problem shows that at a classical level, $\slr$ acts as dynamical symmetry: it changes ${\cal L}$ by total derivative. We  choose the following basis of ${\mathfrak{sl}}(\scri2,\mathbb{R})$, the Lie algebra of $\slr$:
\be
j_0=\left(\begin{array}{cc}
0 & -\frac{1}{2} \\ \frac{1}{2} & 0
\end{array}\right),\,
j_1=\left(\begin{array}{cc}
0 & \frac{1}{2} \\ \frac{1}{2} & 0
\end{array}\right),\,
j_2=\left(\begin{array}{cc}
\frac{1}{2} & 0 \\ 0 & -\frac{1}{2}
\end{array}\right)
\end{equation}
The corresponding constants of motion are obtained through Noether theorem.
Following the canonical quantization procedure one gets
\be
\label{ji}
\begin{array}{lll}
J_0=p_x(x^2-y^2+1)/2+xyp_y-By\\
J_1=p_x(y^2-x^2+1)/2-xyp_y+By\\
J_2=xp_x+yp_y
\end{array}
\end{equation}
where the momentum operators are given by
\be
p_x=-i\frac{\partial}{\partial x},\ p_y=-i\frac{\partial}{\partial y}+\frac{i}{y}
\end{equation}
These generators satisfy the canonical form of the $SO(2,1)$ (or $SU(1,1)$ for the covering group) algebra. This symmetry is of great interest since the Hamiltonian is straightforwardly expressed in terms of the Casimir (note that we use the standard convention for the $SO(2,1)$ norm) $C=||J||^2=J_iJ^i=J_{0}^2-J_{1}^{2}-J_{2}^{2}$:
\be
H=\frac{B^2}{4}-\frac{C}{4}
\end{equation}
which allows to obtain the spectrum. One can explicitly build a standard basis $|jm>$ of the Hilbert space ${\mathfrak{H}}$ such that
\be\label{spectrum}
\begin{array}{lll}
<jm|jm'>&=&\delta_{mm'}\\
C|jm>&=& j(j+1)|jm>\\
J_0|jm>&=&m|jm>
\end{array}
\end{equation}
One distinguishes a continuous and discrete part of the spectrum, depending on the value of $j$. 
In any case, one has 
\be\label{spec}
m=B+n,\ n\in {\mathbb Z} 
\end{equation}

\subsection{Discretization: $B_3$ generators as generalized magnetic translation operators}

We now investigate further the $\slr$ symmetry of this problem,  and in particular we wish  to define generalized magnetic translation  operators, the same way as magnetic translation operators are defined in Euclidean geometry (see \cite{belissard,sunada,shubin} for review). We are looking for a representation of the symmetry group of the problem on the Hilbert space ${\mathfrak{H}}$ of wave functions. For sake of simplicity, we propose a definition in the case of the hyperbolic plane, keeping in mind that a more general formulation follows straightforwardly. The same construction can be performed in particular as soon as the symmetry group is a Lie group; then discretization requires a lattice subgroup of this Lie group.

\vspace{0.8cm}{\bf Definition 3.}  {\it We call generalized magnetic translation operator (GMTO), denoted by ${\cal O}_\tau$, the unitary operator corresponding to an isometry $\tau$ acting on a wave function $\psi$ as follows:
\be
{\cal O}_\tau[\psi](z)=\psi(\tau^{-1}z)
\end{equation}
such that $[{\cal O}_\tau,H]=0$.}

The Wigner theorem (see \cite{sternberg}) shows that such a representation ({\it projective unitary representation}) exists up to a phase:
\be
{\cal O}_{\tau\tau'}=e^{i\Phi(\tau,\tau')}{\cal O}_\tau{\cal O}_{\tau'}
\end{equation}
In particular for a closed sequence of isometries $\disp\gamma=\prod_{i}\tau_i={\rm Id}$, the corresponding product of operators acts on a wave function $\psi$ as follows:
\be
[\prod_{i}{\cal O}_{\tau_{i}}](\psi)(z)=\exp(i\Phi_\gamma)\psi(z)
\end{equation}
Where $\Phi_\gamma$ is the magnetic flux through the loop $\gamma$. 

We obtain explicitly GMTO by exponentiating infinitesimal generators (\ref{ji}). Namely we  consider the following diagram
\be\begin{array}{ccccccc}
\slr & \hfl^{\log}_{} & {\mathfrak{sl}}({\scri2},{\mathbb{R}}) & \hfl^{\phi}_{}& L({\mathfrak{H}}) & \hfl^{\exp}_{} & U({\mathfrak{H}})\\
x & \hfl^{}_{} & \al^i j_i & \hfl^{}_{} & \al^i J_i & \hfl^{}_{} & e^{\disp i\al^i J_i}
\end{array}
\end{equation}
where $L({\mathfrak{H}})$ is the set of linear operators on ${\mathfrak{H}}$ and $U({\mathfrak{H}})$ the group of unitary operators. Then

\vspace{0.8cm}{\bf Proposition 2.}  {\it The following homomorphism

\be\chi=\exp\circ\phi\circ\log:\slr\longrightarrow U({\mathfrak{H}})
\end{equation}
defines a projective unitary representation of $\slr$.}

{\it Proof.} The first ingredients are the usual tools employed to reduce representation problems for  connected  Lie groups to their  Lie algebra: this uses the exponential map and the Campell-Hausdorff formula. Now at the algebra level, note that $\phi$ maps ${\mathfrak{sl}}({\scri2},{\mathbb{R}})$  to a precise representation of the $SU(1,1)$ algebra in $L({\mathfrak{H}})$. Irreducible representations of $SU(1,1)$ have been classified in \cite{bargmann,lindblad}, and the one we consider here is explicitly defined in \cite{alain}. In particular for $B$ non integer, it is a proper representation of the algebra, but a projective representation of the group. This comes from the fact that $SU(1,1)$ is multi-connected.\qed

We now discretize the problem, considering transformations on hyperbolic lattices of ${\cal H}$: we study the representation of $\slz_u$ (which is a lattice subgroup of $\slr$) induced by  $\chi$.  Using the explicit form (\ref{eq:4bis}), we compute the operators associated to elementary steps on $\slz_u$:
\be
\chi(a_u)=\exp(i\pi\al_{u}^iJ_i)\qquad {\rm and}\qquad \chi(b_u)=\exp(\frac{2i\pi}{3}\beta_{u}^iJ_i)
\end{equation}
with $\al_u=\frac{1}{2}\left(u+1/u,1/u-u,0\right);\,\beta_u=\left(2/\sqrt{3},0,-1/\sqrt{3}\right)$. Note that $J$ transforms naturally as a vector of $SO(2,1)$. This is due to the following identity:
\be
[J^{i}u_i,J^{i}v_i]=i\epsilon_{ijk}J^{i}u_{j}v_{k}
\end{equation}
which is just a generalization of the maybe more familiar relations between Euclidean rotations generators ($SU(2)$ case). $\al_u$ (resp. $\beta_u$) being of norm  1, $\al_{u}^iJ_i$ (resp. $\beta_{u}^iJ_i$) has the same spectrum as the compact generator $J_0$ given in (\ref{spectrum}), if one consider a basis obtained by  a nonlinear rotation (depending on $\al_{u}$ or $\beta_{u}$) of the standard basis. Using the spectrum (\ref{spec}),  this allows to compute the phase factors for the elementary loops of $\slz_u$, predicted by the Wigner theorem: 
\be\label{loop}
\chi(a_u)^2=\chi(b_u)^3=e^{2i\pi B}{\rm Id}
\end{equation}
The key point is that this extra relation, introduced by the representation itself, restores the complete algebraic structure of $B_3$. The subgroup of $U({\mathfrak{H}})$ generated by $\chi(a_u)$ and $\chi(b_u)$ is then isomorphic to $B_3$. In other words, we claim that

\vspace{0.8cm}{\bf Theorem 1.}  {\it The homomorphism
\be\Psi=\chi\circ\pi\,:\,B_3\longrightarrow U({\mathfrak{H}})
\end{equation}
defines a  representation of the braid group $B_3$.}

{\it Proof.} It is known that any projective representation of a group is a proper representation of a covering group. Comparing (\ref{loop}) and (\ref{artin}) shows that in our case this covering group is $B_3$.\qed

 Note that $\pi$ and therefore $\Psi$ depend implicitly on the Burau parameter $u$.  If $\theta=2\pi B$,  $\Psi$ coincides with $\Psi_\theta$ introduced previously and defines by extension  a representation of $R_\theta$ (rigorously, the action defined above is on ${\mathfrak{H}}$; we assume that this induces naturally an action on $l^2(\slz)$). $\Psi_\theta(a)$ and $\Psi_\theta(b)$ being by construction GMTO, $\Psi_\theta(h)$ is the discrete Hamiltonian of the problem.

{\it Remark.} The 2--cocycle $\Theta$ is not needed explicitly in this construction. It can nevertheless be obtained naturally as a holonomy factor as in \cite{mathai}.

\section{Derivation of the drift and return probability within this formalism} The point  is that the representation above (we now identify $X$ and $\Psi_{\theta}(X)$), in agreement with proposition \ref{iso1}, allows to define a new decomposition in  $C^{\star}(B_3)$ parametrized by $\slz$. We need to introduce an irreducible form on $\slz$, namely we write each element $\gamma=P_{\gamma}(\pi(a),\pi(b))$, where $P_\gamma$ is a polynomial of minimal order. Then

\vspace{0.8cm}{\bf Proposition 3.}  {\it The following decomposition holds in $C^{\star}(B_3)$}
\be\label{dec}
X=\sum_{\gamma\in\slz}\sum_{n\in\Z}c_{n,\gamma}(X)e^{in\theta}P_{\gamma}(A_\theta,B_\theta)
\end{equation}
This decomposition naturally generates a map $m_{\theta}:C^{\star}(B_3)\longrightarrow C^{\star}(\slz)$ by setting
\be
\begin{array}{ccc}
m_{\theta}(X)&=&\disp\sum_{\gamma\in\slz}\sum_{n\in\Z}c_{n,\gamma}(X)e^{in\theta}\gamma\\
\ &=&\disp\sum_{\gamma\in\slz}{\tilde c}_{\gamma}(m_\theta(X))\gamma
\end{array}
\end{equation}
Note that this map is not a homomorphism, it is built for combinatorial purposes only. We denote by ${\rm tr}$ the tracial state on $C^{\star}(\slz)$, that is the scalar component of an operator: ${\rm tr}(Y)={\tilde c}_{Id}(Y)$. This parameterization leads, after some algebra, to the following:
\be\label{four}
p_n(f)=\frac{1}{2\pi}\int_{-\pi}^{\pi}{\rm tr}(m_\theta(h^n))e^{if\theta}d\theta
\end{equation}
We now use this framework to compute the desired quantities for a random walk on $B_3$. The main results were originally exposed in \cite{voitnech3}, with a more intuitive approach. We here propose a more rigorous proof relying on the connection between $B_3$ and $\slz$ established in theorem 1 and proposition 3. This new approach is also more general since it implies a better understanding of the transition operator $h$ and the algebra $C^{\star}(B_3)$, objects very useful in various contexts. 
\subsection{Return probability and hyperbolic area problem}

\vspace{0.8cm}{\bf Theorem 2.}  {\it The joint probability distribution for a simple random walk on $\slz$ to return to the origin after $n$ steps enclosing an area $f$ is given by:
\be\label{aire} 
p_n(f)\propto\frac{\lambda^n}{n^2}\exp\left(-\frac{f^2}{2n\sigma^2}\right)\end{equation}
$\lambda$ and $\sigma$ depending  on the system of generators. For $S=\{\sigma_1,\sigma_2, \sigma_1^{-1},\sigma_2^{-1}\}$, $\sigma_{S}=1/6$ and $\lambda_{S}=(1+2\sqrt{2})/4$.}

{\it Proof.} Defining the  auxiliary operator
\be
{\tilde h}_{\theta}=\frac{1}{4}(e^{i\theta/2}\pi(a)+e^{-i\theta/2}\pi(a)^{-1}+e^{i\theta/3}\pi(b)+e^{-i\theta/3}\pi(b)^{-1})
\end{equation}
which satisfies ${\rm tr}({\tilde h}^{n}_{\theta})={\rm tr}(m_\theta(h^n))$, allows to derive asymptotics of  $p_n(f)$. We define
\be
{\tilde c}^{n}_k(\theta)=\sum_{\gamma_{k},\,L(\gamma_{k})=k}{\tilde c}_{\gamma_{k}}({\tilde h}_{\theta}^n)
\end{equation}
which  satisfies the following recursion
\begin{eqnarray}
{\tilde c}^{n}_k(\theta)&=&\sum_{\gamma_{k-1},\,L(\gamma_{k-1})=k-1}{\tilde c}_{\gamma_{k-1}}(m_\theta(h^n))g_{1}(\gamma_{k-1})\\
\ & &+\sum_{\gamma_{k},\,L(\gamma_{k})=k}{\tilde c}_{\gamma_{k}}(m_\theta(h^n)g_{0}(\gamma_{k})\\
\ & &+\sum_{\gamma_{k+1},\,L(\gamma_{k+1})=k+1}{\tilde c}_{\gamma_{k+1}}(m_\theta(h^n))g_{-1}(\gamma_{k+1})
\end{eqnarray}
where $g_{i}(\gamma)$ is the factor corresponding to all possible steps $x$ such that $L(\gamma x)=L(\gamma)+i$. A careful counting of the different possible transitions  (such a counting is allowed since we know exactly the irreducible form $P_\gamma$ for $\slz$, see \cite{voitnech3}; only last generator of the irreducible form is important) gives
\be\label{system}
\left\{\begin{array}{ccc}\disp
{\tilde c}^{n+1}_k(\theta)&=&\disp(\frac{\alpha}{2}\cos(\theta/2)+\frac{1-\alpha}{2}\cos(\theta/3))\,{\tilde c}^{n}_{k-1}(\theta)\\
\disp\ &+&\disp\frac{\alpha}{4}\cos(\theta/3)\,{\tilde c}^{n}_k(\theta)\\
\disp\ &+&\disp(\frac{1-\alpha}{2}\cos(\theta/2)+\frac{\alpha}{4}\cos(\theta/3))\,{\tilde c}^{n}_{k+1}(\theta)\ \ \qquad k\ge1\\
\ &\ &\ \\
\disp{\tilde c}^{n+1}_0(\theta)&=&\disp\frac{1-\alpha}{2}\cos(\theta/2)\,{\tilde c}^{n}_{1}(\theta)+\frac{1}{2}\cos(\theta/3)\,{\tilde c}^{n}_0(\theta)
\end{array}
\right.
\end{equation}
where $\alpha$ is a constant to determine, giving the ratio of random words whose irreducible form ends with $b^{\pm1}$. For $\theta=0$ one recovers the master equation governing a random walk on $\slz$, studied throughout \cite{voitnech3}. This allows in particular to compute $\alpha=3/5$. We are interested in the asymptotics $n\gg1$. In this regime, the solution is assumed to be of the form
\be 
{\tilde c}^{n}_k(\theta)=q^{n}_{k}\mu^{n}(\theta)
\end{equation}
This factorization is implicitly accounted for in \cite{voitnech3}: the process generating the phase of the coefficient ${\tilde c}_{\gamma_{k}}(m_\theta(h^n))$, described by $\mu(\theta)$ is independent of  the word length $k$
 in $\slz$ ($q^{n}_{k}$ is, up to normalization, the distribution function of a random walk on $\slz$, computed in \cite{voitnech3}). The system (\ref{system}) then gives asymptotically
\be\label{mu}
\mu^{n}(\theta)=\frac{1}{2^n}(\cos(\theta/2)+\cos(\theta/3))^n
\end{equation}
and we have estimated in particular ${\tilde c}^{n}_0(\theta)={\rm tr}(m_{\theta}(h^n))$. We then apply (\ref{four}) to conclude the proof.\qed

{\it Remark 2.} We could have used instead of $m_\theta$ a $\star$--homomorphism:
\be
M_{\theta}:\left\{
\begin{array}{llll}
A_\theta&\longrightarrow&e^{i\theta/2}\pi(a)\\
B_\theta&\longrightarrow&e^{i\theta/3}\pi(b)
\end{array}\right.
\end{equation}
which gives $M_{\theta}(h)={\tilde h}_{\theta}$. This formalism is equivalent for trace computations, but for other combinatorial matters it can be confusing. In particular the length function is easier to define using $m_\theta$.

Recall that $f$ counts the area enclosed by the random path; it is then interesting to compare this discrete model with  the exact solution of the area problem in the continuous case, derived in \cite{matsumoto}. The probability distribution in this case reads:
\be
p_{t}(A)\propto\frac{\exp(-A^2/2t)}{\cosh^2(\pi A/t)}
\end{equation}
where $A$ is the hyperbolic area and $t$ the time. For long trajectories ($A\gg1$), one can check that the Gaussian behavior (\ref{aire}) of our discrete model is recovered. For short trajectories ($A\ll1$) one recovers the flat space limit, unreachable in our model because the length scale is fixed by the curvature. This asymptotic agreement between the continuous and discrete model is an important fact, though intuitively expectable. Indeed the definition of hyperbolic lattices is not as straightforward as in the Euclidean case (isometric lattices of ${\mathbb R}^n$ are exactly ${\mathbb Z}^n$). In particular the elementary cell of a hyperbolic lattice (or fundamental domain) is neither  necessarily compact nor of finite area (see \cite{terras,elstrodt}); therefore the agreement between the continuous an discrete approach  of the area problem is not trivial, and supports the idea  that $\slz$ is a ``good'' discretization of $\h$.

\subsection{The drift, or average topological complexity}

\vspace{0.8cm}{\bf Theorem 3.} {\it The drift of a simple random walk on $B_3$ is given by:
\be
l_{B_3}=\lim_{n\to\infty}\frac{\left<L^{B_3}(w_n)\right>}{n}=\frac{1}{4}\end{equation}}
{\it Proof.} We use (\ref{drift}) and the decomposition (\ref{dec}) to write 
\be
l_{B_3}=\lim_{n\to\infty}\frac{1}{n}\sum_{\gamma\in\slz}\sum_{k\in\Z}c_{k,\gamma}(h^n)L^{B_3}(e^{ik\theta}P_{\gamma}(A_{\theta},B_{\theta}))\end{equation}
Using $L^{\slz}(P_{\gamma}(A_{\theta},B_{\theta}))\le L^{B_3}(e^{ik\theta}P_{\gamma}(A_{\theta},B_{\theta}))\le L^{\slz}(P_{\gamma}(A_{\theta},B_{\theta}))+|k|$, we first show that 
\be\label{bound}
\frac{L(\pi(w_n))}{n}\le \frac{L^{B_3}(w_n)}{n}\le\frac{L(\pi(w_n))}{n}+\frac{1}{n}\sum_{k\in\Z}|k|c_{k,w_{n}}(h^n)
\end{equation}
The last ingredient is the following inequality: 
\be
\sum_{k\in\Z}|k|c_{k,\gamma}(h^n)\le |\frac{d}{d\theta}{\tilde c}_{\gamma}({\tilde h}_{\theta})|.\end{equation}
Using (\ref{mu}), the following holds
\be\label{bound2}
\frac{\left<L(\pi(w_n))\right>}{n}\le \frac{\left<L^{B_3}(w_n)\right>}{n}\le\frac{\left<L(\pi(w_n))\right>}{n}+O\left(\frac{1}{\sqrt{n}}\right)
\end{equation}
We use the  result $\disp l_{\slz}=1/4$ obtained 
in \cite{voitnech3} for symmetric random walk on $\slz$ to conclude the proof.\qed

A more intuitive approach  of this result is proposed in \cite{voitnech3}. The development carried out above shows that this result could  be generalized to many other groups: if $E$ is an Abelian extension of a hyperbolic group $G$, then the drift on $E$ equals the drift on $G$. We here give this result without proof.  

\section{Conclusion and Remarks}
To conclude, we wish to comment the new physical content of these results. So far the only topological invariant studied in the context of entanglements of fluctuating linear objects was the so-called winding number \cite{belisle,alain1,nelson,kardar}, which is known to be incomplete for more than 2 linear objects. We here propose a conceptually simple model (in particular fluctuations are not precisely taken into account), which describes exactly the underlying non-Abelian topology of the problem. The resulting behavior of the topological  complexity, the irreducible braid length,  differs strongly from the Abelian case. Indeed, for $n$ elementary moves, the winding number $W$ for our three strands problem would scale like $\sqrt{\left<W^2\right>}\propto\sqrt{n}$ (compare with $\left<L(w_n)\right>\propto n$), whereas the probability of $W=0$ would scale like $n^{-3/2}$ (compare with $p_{n}(0)\propto\lambda^n/n^2$).     
 
 Beside these direct motivations of the physical model, this work allowed to study the link, through the example of $B_3$,  between random walks on a discrete group and a projective unitary representation of this group in a Hilbert space of a quantum system in a magnetic field. This is in our case the hyperbolic counter part of the well--known \ho problem in Euclidean geometry. We could in particular reformulate and asymptotically compare the distribution of the hyperbolic area enclosed by a random path, in the case of a lattice random walk  and in the case of a continuous Brownian motion. This correspondence requires to define generalized magnetic translation operators, and in particular leads to a general algebraic set up for discrete problems with magnetic fields, which enabled us to rigorously prove different results concerning $B_3$ already obtained in \cite{voitnech3}. We hope that this extension of the problem to hyperbolic geometry is not restrictive and could be adapted to other spaces.

{\bf Acknowledgements}

It is a pleasure to thank A. Comtet, P. Dehornoy and  S. Nechaev  for stimulating discussions and useful comments.

\end{document}